\documentclass[12pt]{statsoc}

\usepackage[margin=1in]{geometry}

\usepackage{enumitem}   
\usepackage[T1]{fontenc}
\usepackage[utf8]{inputenc}
\usepackage{amsmath,amssymb}
\usepackage{pgfplots}

\usepackage{array}
\usepackage{calc}
\usepackage{wrapfig}

\usepackage{color, soul}
\usepackage{cleveref}
\usepackage{subfigure}
\usepackage{booktabs}

\title{Towards more scientific meta-analyses}
\author[Lily H. Zhang, Menelaos Konstantinidis, Marie-Abèle Bind, Donald B. Rubin]{Lily H. Zhang\textsuperscript{1}, Menelaos Konstantinidis\textsuperscript{2},
Marie-Abèle Bind\textsuperscript{3,4}, 
Donald B. Rubin\textsuperscript{5,6,7}}
\address{\textsuperscript{1}Center for Data Science, New York University}
\address{\textsuperscript{2}Li Ka Shing Knowledge Institute, St. Michael's Hospital, Unity Health Toronto, ON, Canada}
\address{\textsuperscript{3}Biostatistics Center, Massachusetts General Hospital, Boston, MA 02114}
\address{\textsuperscript{4}Department of Medicine, Harvard Medical School, Boston, MA 02115}
\address{\textsuperscript{5}Professor Emeritus, Department of Statistics, Harvard University, \\Cambridge MA, US}
\address{\textsuperscript{6}Professor, Yau Center for Mathematical Sciences, Tsinghua University, Beijing China}
\address{\textsuperscript{7}Murray Shusterman Senior Research Fellow, Department of Statistical Science, Fox School of Business, Temple University, Philadelphia, PA}

\pgfplotsset{compat=1.15}
\begin{document}
\begin{abstract}
Meta-analysis can be a critical part of the research process, often serving as the primary analysis on which the practitioners, policymakers, and individuals base their decisions. However, current literature synthesis approaches to meta-analysis typically estimate a different quantity than what is implicitly intended; concretely, standard approaches estimate the average effect of a treatment for a population of imperfect studies, rather than the true scientific effect that would be measured in a population of hypothetical perfect studies. We advocate for an alternative method, called \textit{response-surface meta-analysis}, which models the relationship between the quality of the study design as predictor variables and its reported estimated effect size as the outcome variable in order to estimate the effect size obtained by the hypothetical ideal study. The idea was first introduced by Rubin several decades ago \cite{Rubin}, and here we provide a practical implementation. First, we reintroduce the idea of response-surface meta-analysis, highlighting its focus on a scientifically-motivated estimand while proposing a straightforward implementation. Then we compare the approach to traditional meta-analysis techniques used in practice. We then implement response-surface meta-analysis and contrast its results with existing literature-synthesis approaches on both simulated data and a real-world example published by the Cochrane Collaboration. We conclude by detailing the primary challenges in the implementation of response-surface meta-analysis and offer some suggestions to tackle these challenges.
\end{abstract}

\section{Background}
Gene Glass first coined the term ``meta-analysis'' in 1976 to mean ``the statistical analysis of a large collection of individual studies'' \cite{Glass}. 
Many statistical methods have emerged to address various challenges in conducting a meta-analysis, 
from various estimators for between-study heterogeneity \cite{Veroniki} to methods for detecting and possibly correcting publication bias \cite{Thornton}.
Yet the central methodology in a meta analysis typically involves combining studies via some form of averaging, such as is the case when estimating the relevant mean under a fixed-effect(s) or random-effects model \cite{Deeks}. Consequently, there is an implicit assumption that the researcher's target is a summary of the existing literature. 

With current meta-analysis methodologies, which we will refer to as ``literature synthesis meta-analysis,'' we are addressing the question, ``What is the typical effect in the underlying population of studies from which the published studies can be viewed as a random sample?'' Although this is a reasonable question, it is usually not the actual question researchers set out to address when they conduct a meta-analysis. Instead, the hope is usually to use the collection of existing research studies to better estimate the true scientific effect if it could be measured \textit{perfectly}. The latter is the focus of response-surface meta-analysis, which has been praised by the inventor of the concept of meta-analysis himself \cite{Glass review}.

In this work, we formulate response-surface meta-analysis in \Cref{sec:background} and compare it with existing approaches in \Cref{sec:related}. Then, we show how response-surface meta-analysis can yield different conclusions from traditional approaches via a simulation study where the true effect is known (\Cref{sec:simulated}) as well as via a real-world example published by the Cochrane Collaboration (\Cref{sec:real_world}).

\section{Formulating Response-surface Extrapolation}
\label{sec:background}
Here, we review the central parts of the response-surface approach first proposed in \cite{Rubin}. Let us suppose that the goal is to estimate the true effect of a treatment relative to a control on an outcome. We denote this effect as $\tau$. Individual studies report average effect sizes 
$\psi$,
which can be influenced by factors $(X, Z)$; here, $X$ denotes scientific factors, or variables that have scientific interest, such as sex, age, or education of study participants; Z denotes design factors that influence the quality of studies, such as sample size and the experimental design itself (e.g., randomized vs. observational). 
The response-surface is a probabilistic model for 
$\psi$
as a function of $X$ and $Z$, denoting how both scientific and design factors can influence the study effect sizes reported by individual studies. To estimate the true effect size $\tau_x$ for given scientific factors $X=x$, we need to estimate this response-surface where $Z$ is extrapolated to the perfect study, which we denote as $z_\infty$. The reason for this focus is that scientifically, we are only interested in the effect as a function of scientific factors $X$ in the collection of perfect studies.

The primary benefit for response-surface approach is its theoretical clarity and focus on the \textit{correct estimand}.
For further discussion, see Rubin 1990 \cite{Rubin}. 

In this work, we implement response-surface meta-analysis using fixed- and random-effects meta-regression \cite{HigginsMR} with study design as a covariate. Then, we report the predicted mean effect under a prespecified ideal study design. This approach allows us to take advantage of the statistical advances in meta-analysis research made since response-surface meta-analysis was first proposed \cite{Tipton}.

\section{Comparing Response-Surface and Traditional Meta-Analysis}
\label{sec:related}
The primary difference between literature synthesis and response-surface meta-analysis is the estimand of interest. Traditionally, the conceptual framework for meta-analysis assumes that the study-specific effect sizes are distributed around some mean treatment effect. 
However, estimating the mean treatment effect for an underlying population of existing imperfect studies is not the same as estimating the mean treatment effect that would be observed in a population of perfect studies. In other words, current literature-synthesis meta-analysis methods focus on the estimands at each value of X=x, $\tau_{\text{synthesis},x} = \mathbb{E}_\psi(\psi|X=x) = \mathbb{E}_{\psi}\mathbb{E}_Z(\psi|X=x,Z) $, where the distribution of $Z$ is that of the population of study designs of which published studies are a random sample. In contrast, response-surface meta-analysis targets the estimand $\tau_{\infty,x} = \mathbb{E}_\psi(\psi|X=x, Z=z_\infty)$, 
which we argue
is more conceptually aligned with the scientific purpose of meta-analysis.

Literature synthesis meta-analysis takes into account study design but in a different way than response-surface meta-analysis does.
Namely, literature synthesis methods will often bias the distribution of study design quality towards higher quality studies by excluding certain ``unworthy'' studies in the analysis. Examples include  considering only randomized control trials or peer-reviewed articles in an analysis.
However, averaging over such studies is still distinct from extrapolating to the ideal study; by excluding studies, one has changed the distribution of studies being considered, but the included studies themselves do not typically reflect a sample of perfect studies.
In addition, literature-synthesis methods choose a hard threshold for study inclusion and exclusion whereas response-surface meta-analysis  
takes advantage of all studies to fit the response-surface and estimate the effect under perfect conditions as well as possible. 

Our implementation of response-surface meta-analysis is closely related to
meta-regression,
the method of identifying sources of heterogeneity in a meta-analysis. A meta-regression regresses study effects on covariates such as study design or scientific factors to estimate whether such factors significantly affect the measured effect. In this paper, we implement response-surface meta-analysis by first performing a meta-regression using design quality as the independent variable and then using this regression model to estimate the predicted mean effect under optimal study design conditions. How the resulting model is used for analysis is where meta-regression and our implementation of response-surface meta-analysis differ: meta-regression focuses on interpreting the coefficients in the regression (e.g., testing significance), whereas response-surface meta-analysis focuses on the predicted mean effect conditioned on perfect study quality. Response-surface meta-analysis is not concerned with the significance (or lack thereof) of an associated coefficient; instead, any uncertainty associated with the coefficient estimates, whether they are significant or not, is reflected in the resulting standard errors of the estimate of the predicted mean at $X=x$.

\section{Case Study with Simulated Data}
\label{sec:simulated}
Here, we highlight the differences between response-surface and literature-synthesis meta-analysis in 
a simulated data setup. Consider a situation where, as the studies get better, the treatment effect gets more pronounced. Examples include effects that can only be estimated as scientific instrumentation and measurement improve, or interventions which only exhibit an effect once the necessary measures to ensure increased compliance are put in place. 
\subsection{Setup}
\begin{figure}
    \centering
    \subfigure[Design quality.]{\includegraphics[width=55mm]{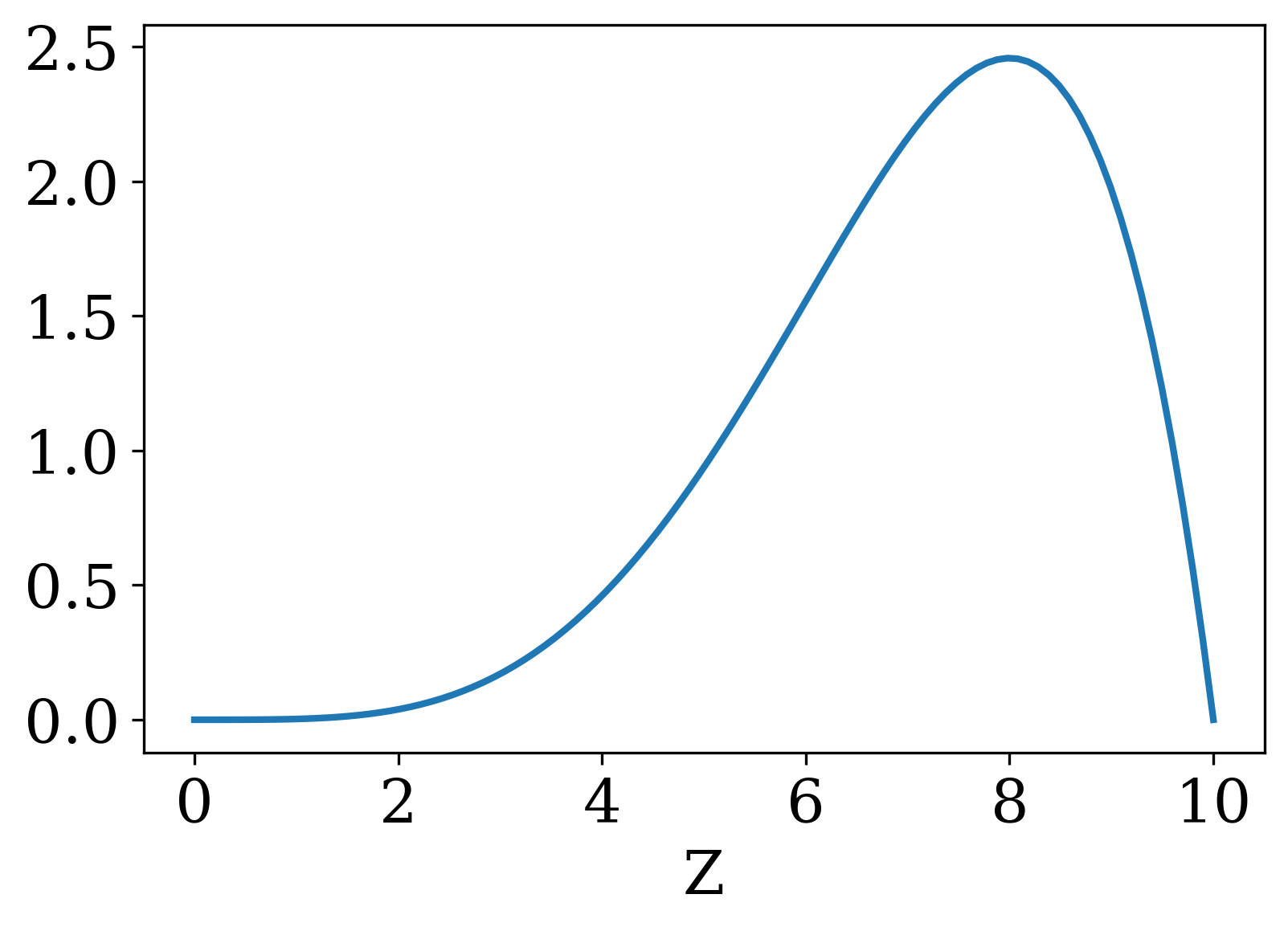}}
    \subfigure[Individual studies.]{\includegraphics[width=59mm]{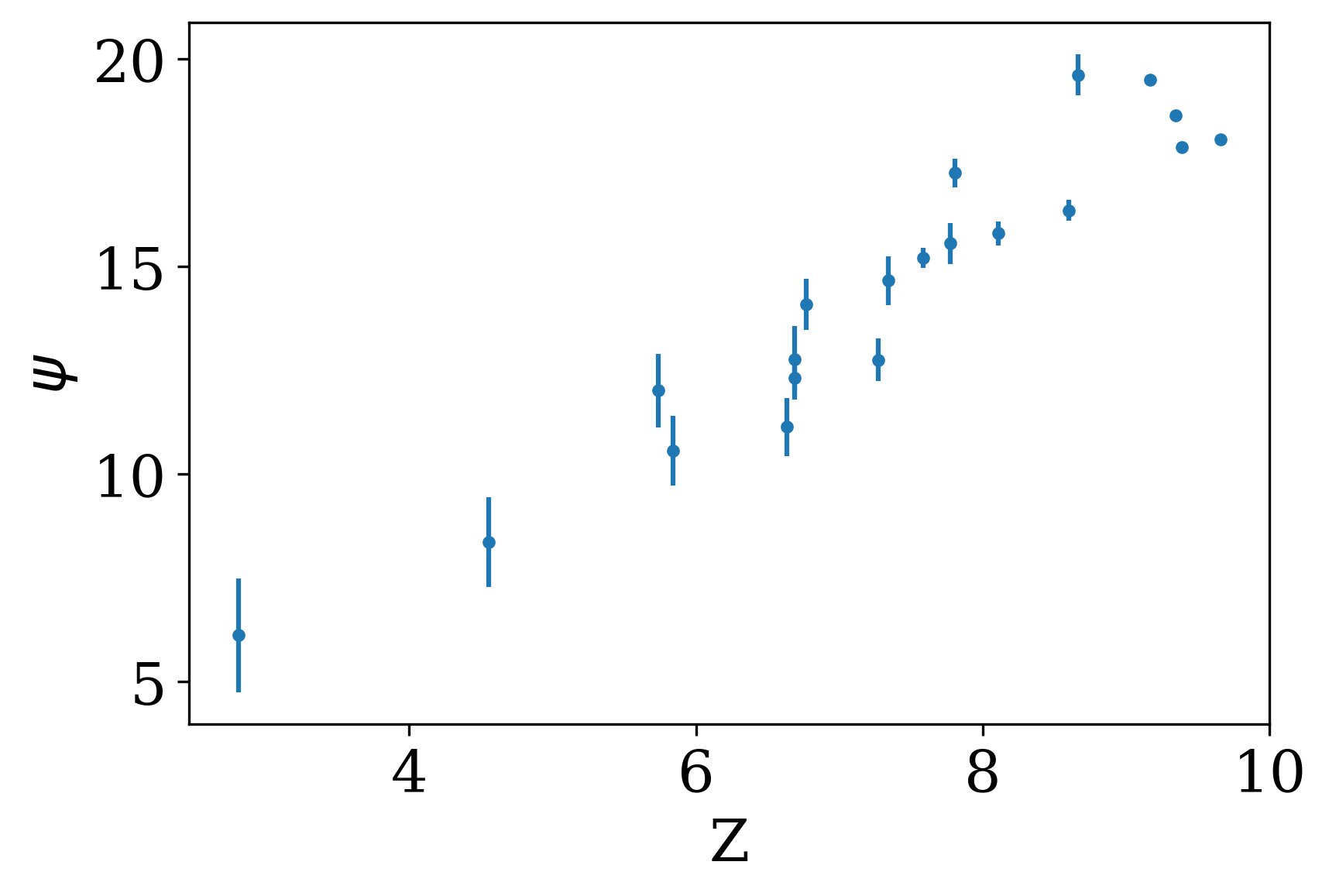}}
    \caption{Simulation study setup. Study design qualities are sampled from the distribution $\text{Beta}(5, 2)$ in (a), and individual study estimated mean effects $\psi$ and standard errors $s$ are sampled subsequently, resulting in (b).}
    \label{fig:z}
\end{figure}

In our simulation, we let $Z$ be a univariate measure of design quality, where $z_\infty = 10$ defines the best design quality possible. We assume that existing studies have design qualities $Z$ that follow a scaled Beta distribution: $Z \sim 10 \; \text{Beta}(\alpha=5,\beta=2)$. Studies can therefore range in quality between 0 and 10, and most studies have reasonably high design quality (i.e., mode at 8). See \Cref{fig:z} (a) for the sampling distribution for design quality.

Next, we sample $n=20$ studies in the following way. First, we sample a design quality score $Z_i$ from the aforementioned sampling distribution. Then, we sample a sample mean and standard error reported by each study as a function of the design quality of the study. For a given study, we draw a sample mean effect $\psi_i$ assuming that individual studies report sample means that are on average closer to the true effect $\tau$ when the design quality is higher: $\psi_i  \sim \mathcal{N}(\tau - \gamma(Z_{\infty} - Z_i), 1)$, where $\gamma=2$. Then, we draw individual standard errors $s_i$ for each study assuming that standard errors of a study's estimate shrink with better design quality: $s_i = \max \Big(\delta (Z_{\infty} - Z) + \epsilon_i, 0.1 \Big)$, where $\delta = 0.2$ and $\epsilon_i \sim \mathcal{N}(0, 0.01)$. 

\subsection{Methods}
We now compare estimated effect sizes using literature-synthesis and response-surface meta-analysis approaches.
For our literature-synthesis meta-analysis, we provide results under both fixed-effects and random-effects models, the latter using the Dersimonian-Laird method. Following convention, we report the estimated mean treatment effect and its 95\% confidence interval ($\pm 1.96 \; \text{SE}$).
For response-surface meta-analysis, we implement both a fixed- and random-effects-based response-surface. For the former,
we fit a fixed-effects meta-regression (i.e., weighted least squares regression with inverse study variances as weights) with design quality as a covariate. Then, we report the predicted mean treatment effect conditioned on ideal design quality (i.e., $Z = 10$). We do the same for the random-effects-based response-surface, except that we fit a random effects meta-regression using the Dersimonian-Laird method to estimate the between-study variance. We report the 95\% confidence interval of both estimates.

\begin{figure}
    \centering
    \includegraphics[width=90mm]{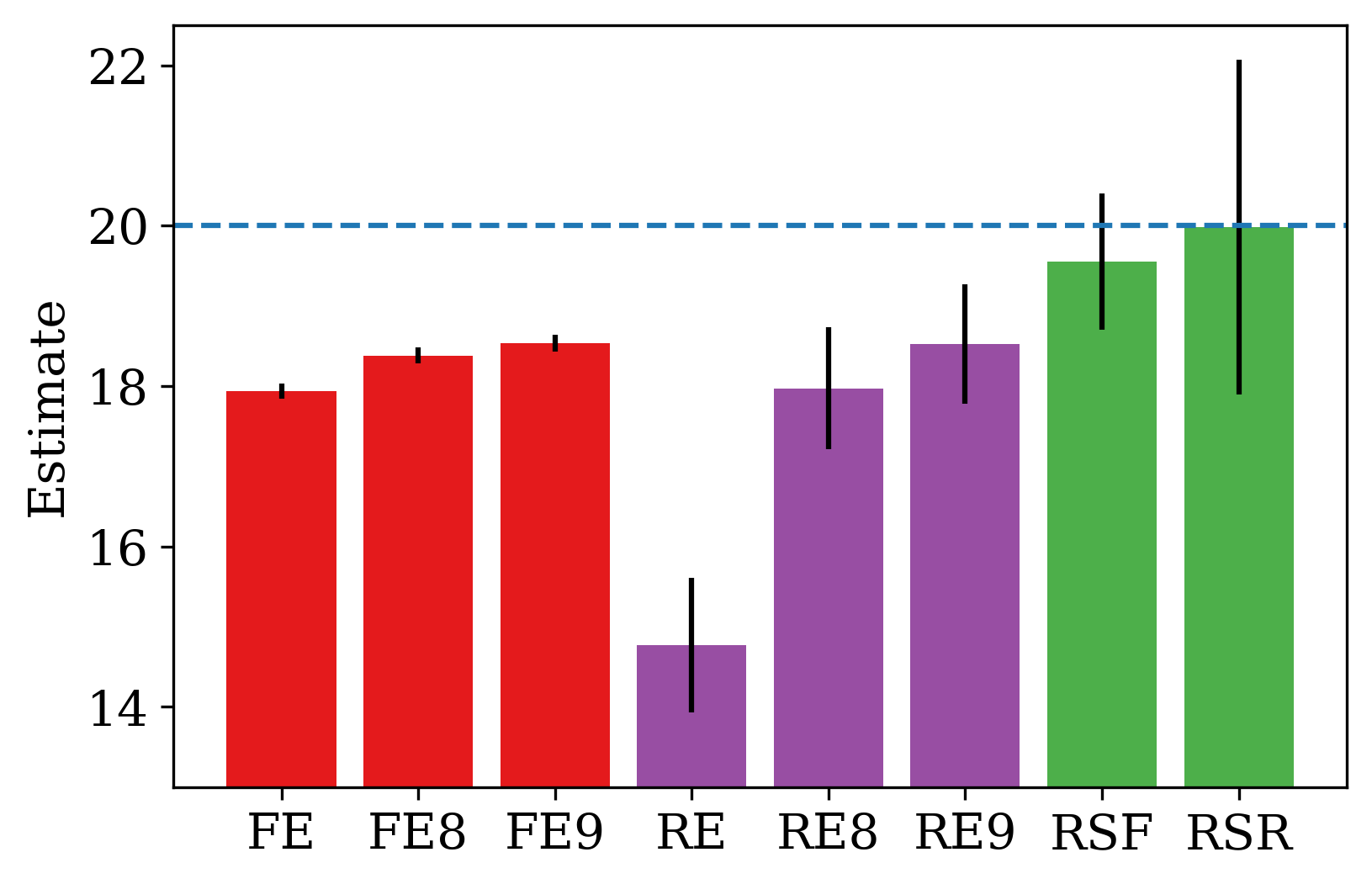}
    \caption{A comparison of literature synthesis (red, purple) and response-surface (green) meta-analysis on the simulated setup in \Cref{fig:simulated_results}. The dashed line indicates the true scientific effect, which only the response-surface approaches (RSF, RSR) capture in their 95\% confidence interval. Fixed effects (FE) and random effects (RE) meta-analysis underestimate the true scientific effect even when the analysis only includes studies with high design qualities (e.g., FE8 and RE8 only include studies with $Z > 8$).}
    \label{fig:simulated_results}
\end{figure}
\subsection{Results}
\Cref{fig:simulated_results} presents the differences in results (estimate and 95\% confidence interval) between the literature-synthesis and response-surface approaches.
In comparison to the true scientific effect size $\tau = 20$, the fixed-effects meta-analysis (FE), random-effects meta-analysis (RE), and response-surface models (RSF for fixed-effects version and RSR for random-effects version) yield estimates of 17.9 (95\% CI: 17.8, 18.0), 14.8 (95\% CI: 13.9, 15.6), 19.6 (95\% CI: 18.4, 20.7), and 20.0 (95\% CI: 17.9, 22.0),
respectively.
In other words, the literature synthesis approaches produce estimates far from the true effect size $\tau$, whereas 
the response-surface estimates are close to the true $\tau$ (within one standard error).

When only considering high quality studies (i.e., those with design qualities of 9 and above), the fixed-effects and random-effects models yield estimates that are closer but still do not cover the true effect size: the fixed-effects estimate is 
18.5 (95\% CI: 18.4, 18.6),
and the random-effects estimate is 
18.5 (95\% CI: 17.8, 19.3).
Including only high-quality studies brings the literature synthesis results closer to the true scientific effect, but these methods still yield biased estimates of the scientific estimand of interest since neither appropriately address the relationship between design quality and effect size. In contrast, given an accurate model of the relationship between design quality and study results, response-surface meta-analysis can appropriately estimate the average effect size conditioned on ideal design quality. 

It is worth noting that the response-surface estimates exhibit wider confidence intervals than the corresponding literature synthesis approaches. This occurs because the latter assume a fixed relationship between design quality and reported effect (i.e., none) whereas the former relaxes this assumption and instead estimates this relationship, the estimate of which incorporates additional uncertainty into the estimate for $\tau$. Thus, when there is indeed no relationship between design quality and individual study estimates, literature-synthesis approaches should provide accurate estimates with tighter confidence intervals than response-surface approaches.
However, the absence of a relationship is unlikely to be true in practice, as various study design decisions (e.g., sample size, blinding) generally affect the bias and variance of an individual study's effect estimate. 

To highlight the differences between literature synthesis and response-surface approaches further, we run a simulation studying the same setup described above while varying the parameters of the data generating process. We allow the following parameters to vary: 1) distribution of design quality $Z$ via the parameter $\alpha$; 2) $\gamma$, the unit change in the bias of an observed study estimate relative to the true estimate $\tau$ for every unit change in design quality; and 3) $\delta$, the unit change in the standard error of the observed study estimate for every unit change in design quality.
We consider four $\alpha$ values (i.e., $\alpha$=2, 5, 8, 11), where increasing $\alpha$ results in a distribution where a larger proportion of studies are of high-quality: $\alpha=2$ yields a symmetric Beta distribution with mode at 5 whereas $\alpha=11$ yields a distribution where over 65\% of the studies have a quality score about 9. We choose four $\gamma$ values (i.e., $\gamma$=0, 1, 2, 4), where 0 suggests that the center of the distribution of study effect means is the same regardless of study design. 
As the value of $\gamma$ increases, lower-quality studies on average exhibit effects $\psi$ that deviate farther from the true scientific effect $\tau$.
Finally, we choose $\delta$ values (i.e., $\delta$=0, 0.1, 0.2, 0.4, 0.8), where 0 suggests that on average, all studies report the same standard errors regardless of design quality. 
As the value of $\delta$ increases, so do the standard errors associated with lower-quality studies.
We vary all factors fully factorially and run 100 repetitions for each setting.

In the following plots, we compare methods based on their median bias and coverage (i.e., proportion of runs where the 95\% CI covers the true effect $\tau$). We compare response-surface meta-analysis (i.e., RSF, RSR), described above, with literature synthesis-based fixed-effects (FE) and random-effects (RE) meta-analysis, as well as fixed-effects and random-effects meta-analysis using only studies with design quality higher than 9 (i.e., FE9, RE9). We see that results are quite consistent across the values of sample size $N$ we examined (i.e., N=10, 20, 40, 80) and report results at $N=20$ to stay consistent with the above case study.

\begin{figure}
    \centering
    \includegraphics[width=1\linewidth]{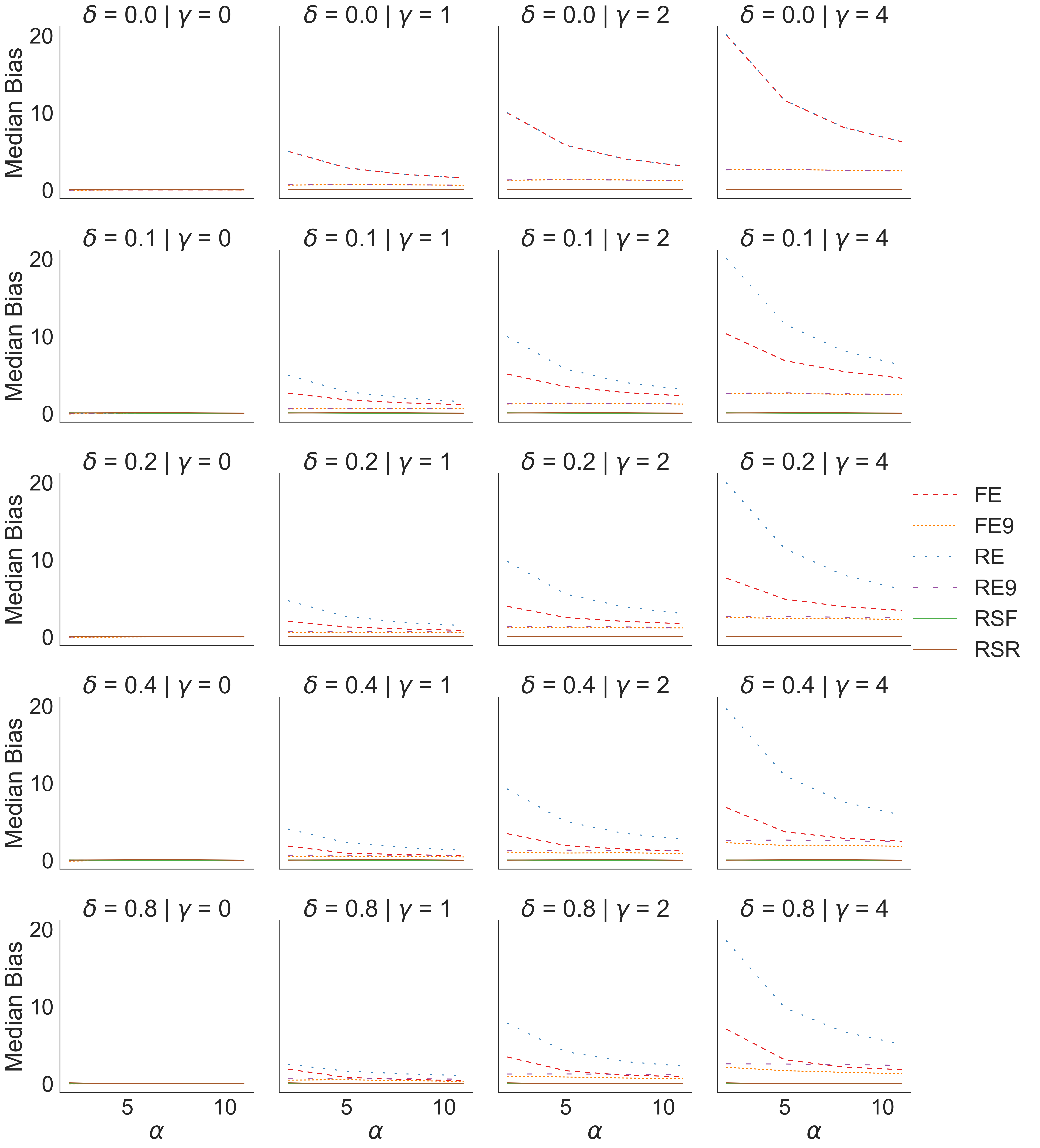}
    \caption{Median bias of literature-synthesis approaches (i.e., FE, FE9, RE, and RE9) can be quite large for various data generating processes, while the median bias for response-surface approaches (i.e., RSF and RSR) stays near-zero for all simulation settings. As the true effect $\tau$ is fixed at 20 for all settings, results from this simulation suggest that literature-synthesis approaches can yield biases as large as magnitude of the true effect itself. }
    \label{fig:median_bias}
\end{figure}

\begin{figure}
    \centering
    \includegraphics[width=1\linewidth]{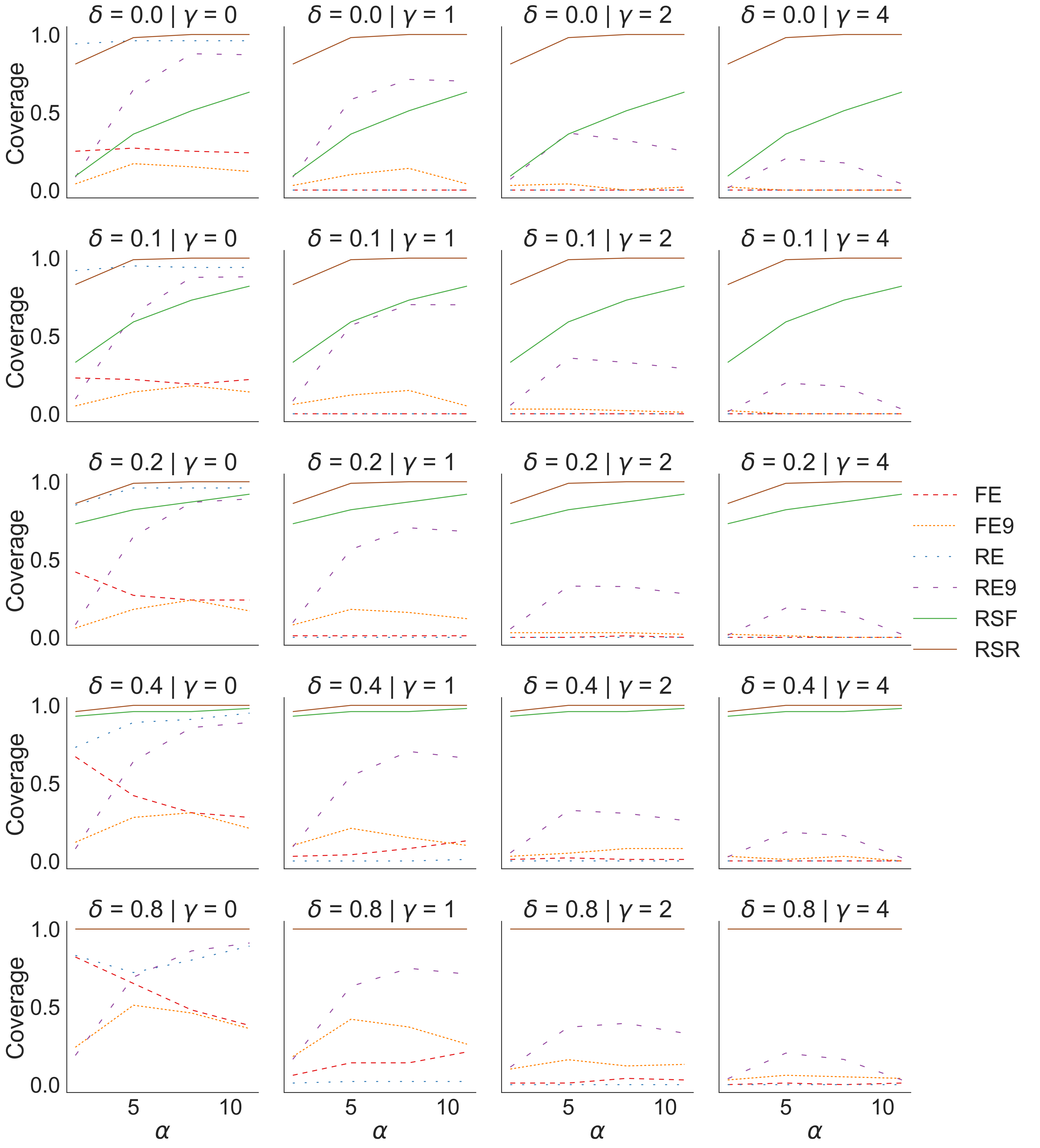}
    \caption{The coverage (i.e., percentage of 95\% confidence intervals that cover the true effect) stays high for response-surface approaches (i.e., RSF, RSR) but varies widely for literature-synthesis approaches (i.e., FE, FE9, RE, and RE9).}
    \label{fig:coverage}
\end{figure}

\Cref{fig:median_bias} shows that the median bias of literature synthesis-based estimates (i.e., FE, FE9, RE, and RE9) increases as $\gamma$ increases (i.e., larger deviations in individual reported effects), where the bias is larger in FE and RE as $\alpha$ decreases (i.e., more lower-quality studies). In contrast, response-surface approaches (i.e., RSF and RSR) maintain a near-zero median bias regardless of the values of $\alpha, \gamma, \delta$. \Cref{fig:coverage} shows that response-surface meta-analysis almost always has the highest coverage, especially when comparing the response-surface approach with the analogous literature-synthesis approach (i.e., RSF vs. FE, RSR vs. RE). The only time literature-synthesis approaches exhibit similar or better coverage than their analogous response-surface approaches occurs for the lowest values of $\alpha, \gamma, \delta$ (i.e. top-left corner of \Cref{fig:coverage}). In other words, literature synthesis meta-analysis is advantageous only when there truly is no relationship between design quality and reported study effect. Otherwise, literature synthesis approaches can yield significant bias and sometimes as low as near-zero coverage.

\section{A Real Data Example}
\label{sec:real_world}
We now present an application of the response-surface approach to real data from a meta-analysis published in the Cochrane Database of Systematic Reviews.

\subsection{Setup}
We re-analyze a meta-analysis comparing the joint effect of diet and physical activity on the z-score of BMI (zBMI) among children aged 6 to 12 years old \cite{Brown19}.
The original meta-analysis consists of 25 studies, all randomized control trials.

Although the definition of the perfect study has not been established in general, to motivate the use of the response-surface methodology, we assign design quality using the Cochrane Risk of Bias (RoB2) tool \cite{HigginsBias}, which provides an ordinal rating (i.e., low, unclear, and high risk of bias) for each of the following six categories: 
    (i) bias due to random sequence generation, which occurs when the process used to generate a random allocation sequence for assigning participants to different groups in a study is flawed or inadequate, leading to an unequal distribution of known or unknown confounding factors between the groups;
    (ii) bias due to allocation concealment, or the possibility that knowledge of the allocation sequence could influence the allocation of participants to different groups in a study, leading to a biased distribution of known or unknown confounding factors between the groups;
    (iii) bias due to blinding, or the possibility that knowledge of the treatment or intervention being received by participants or researchers could influence the measurement or reporting of outcomes;
    (iv) bias due to incomplete outcome data, which can occur when participants drop out of a study or are lost in follow-up;
    (v) bias due to selective reporting, which occurs when the study results and analyses are reported selectively based on the direction or statistical significance of the findings; and
    (vi) other bias.
Based on these ratings, individual studies are assigned either a low, unclear, or high risk of bias overall. 

\begin{figure}
    \centering
    \subfigure[Forest plot of studies measuring effect of diet and exercise on zBMI for children ages 6-12.]{\includegraphics[trim=0 0 -12mm 0, width=110mm]{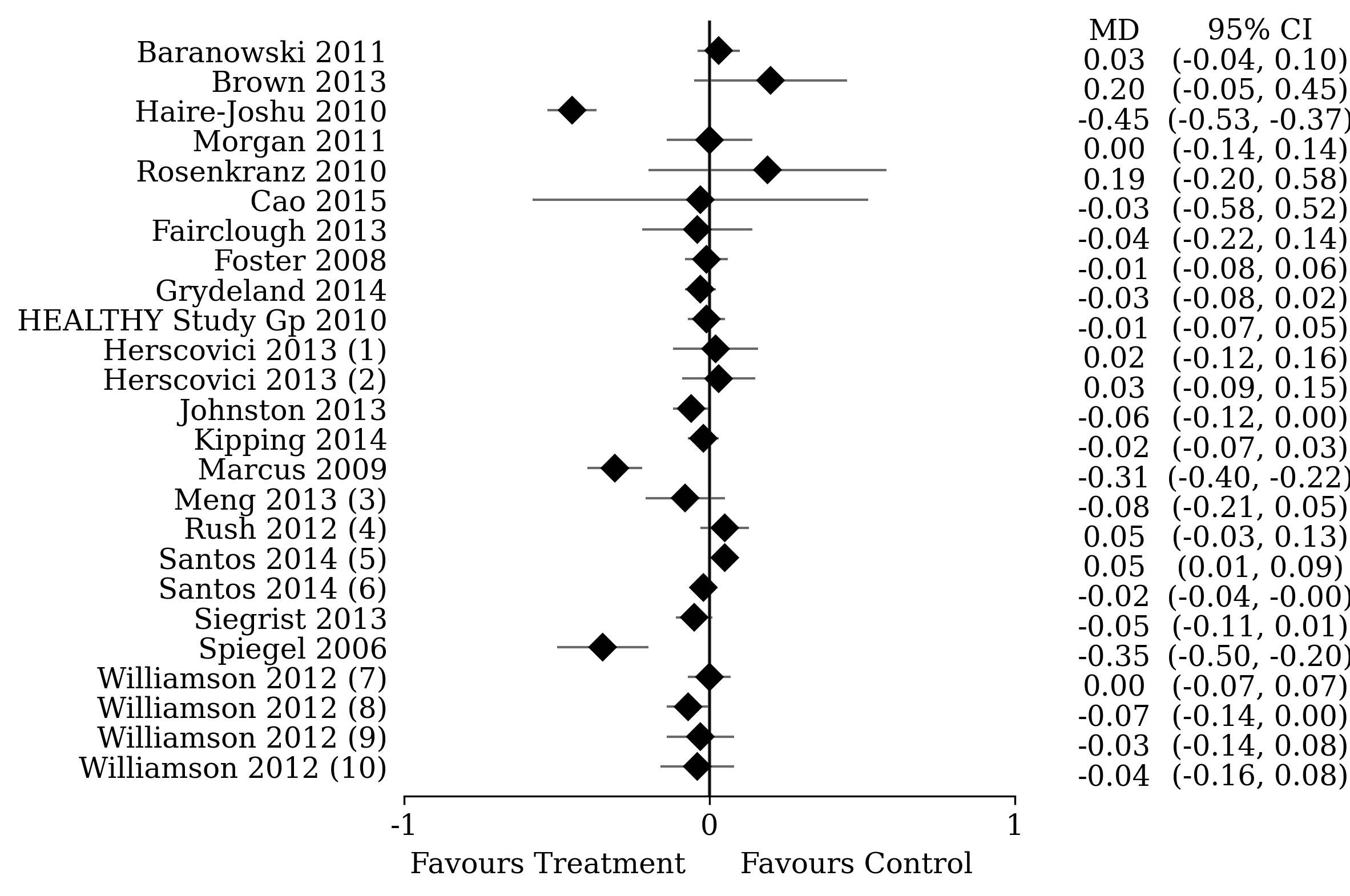}}
    \subfigure[Study effect sizes categorized by risk of bias.]{\includegraphics[trim=0 0 -5mm 0, width=110mm]{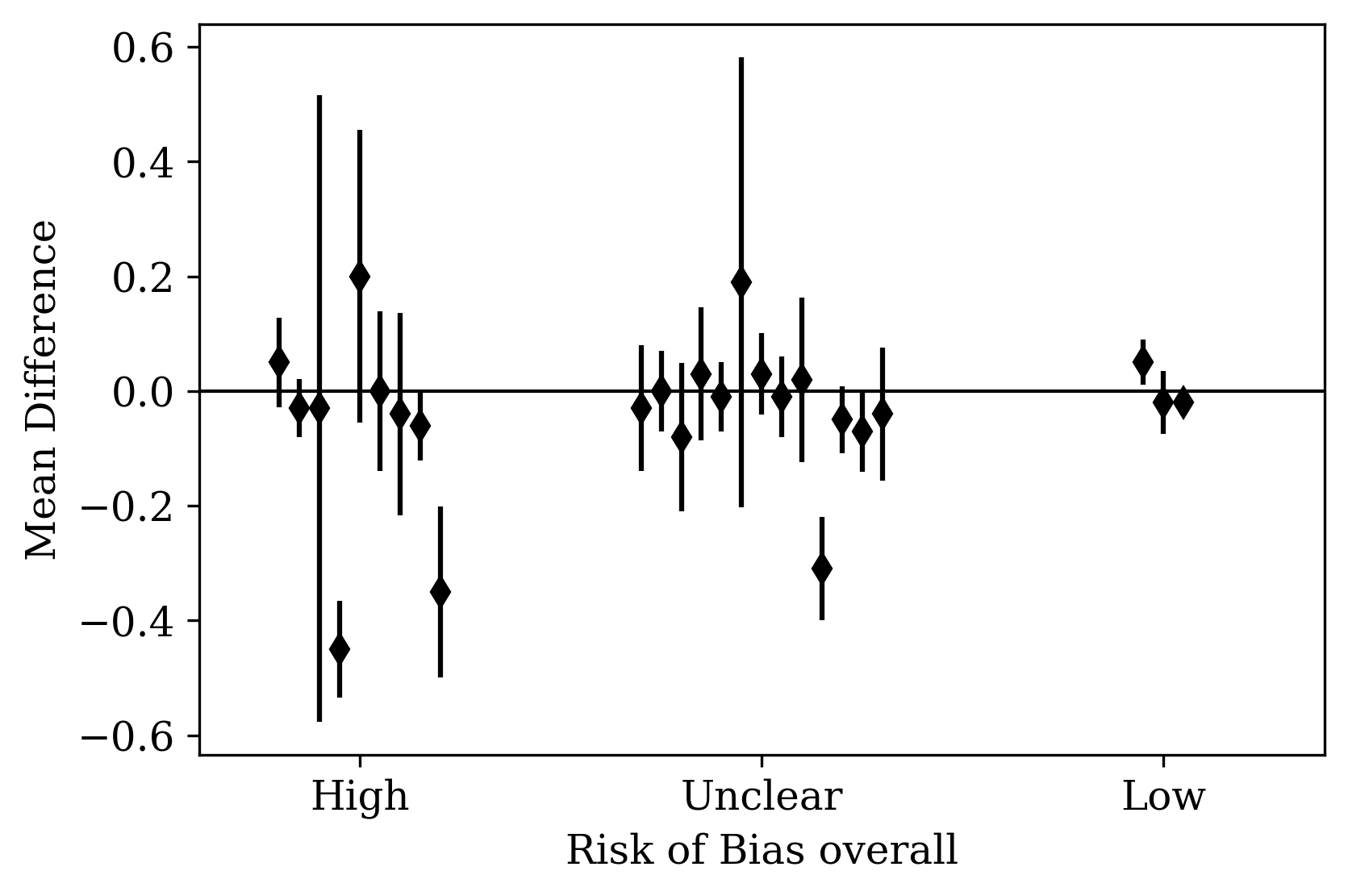}}
    \caption{(a): A forest plot of the results only takes the study estimates and their 95\% CIs into account. The corresponding random effects model estimate favors treatment with estimate -0.05 (CI: -0.10, -0.01). (b): Plotting the study estimates and CIs against overall risk of bias reveals that studies with higher risk of bias tend to favor the treatment (negative mean difference) as well as exhibit wider CIs. The response-surface estimate concludes not enough evidence of an effect, with estimate 0.01 (CI: -0.17, 0.19).}
    \label{fig:real_data_setup}
\end{figure}

We define the ideal study quality (i.e., $z_\infty$) to be a study with low risk of bias overall, which only three of the 25 studies satisfy. Thirteen studies have an unclear risk of bias, and nine have a high risk of bias. To visualize the additional information present when considering study results in relation to design quality, see \Cref{fig:real_data_setup} (b) for an visualization of the study effect estimates categorized by overall risk of bias, as compared to a forest plot of the studies in \Cref{fig:real_data_setup} (a).

\subsection{Methods}
To perform response-surface meta-analysis, we model the average study effect with a random effects meta-regression that includes risk of bias as an ordinal covariate. 
To estimate $\tau$, we report both the mean of the prediction at a low risk of bias, along with its 95\% confidence interval. We compare our result to that of \cite{Brown19}, who utilize a random effects model without taking risk of bias into account.

\subsection{Results}
In contrast to the negative effect estimate published by Brown et al. of
-0.05 (95\% CI: -0.10,-0.01) favoring the intervention, 
the response-surface approach yields a non-negative estimate 
of 0.01 (95\% CI: -0.17, 0.19).
Thus, by adjusting for risk of bias and estimating the effect under ideal study conditions rather than averaged over existing published studies, the inference regarding the treatment effect changes from favoring the intervention to concluding there is not enough information to reject the null hypothesis. 

The forest plot and mean difference vs. risk of bias plots in \Cref{fig:real_data_setup} help build intuition about this result: the forest plot shows that there exist several studies in the set of published studies that show a significant effect favoring the treatment, such that performing a weighted averaging of all studies via the assumptions of a random effects model yields an overall negative effect. However, the corresponding mean difference vs. risk of bias plot shows that the studies that report the most negative effects also have high or unclear risk of bias, while the few low risk of bias studies yield precise effect estimates close to zero mean difference. 
Rather than allowing the higher risk of bias studies to bias the overall average towards a negative effect,
the response-surface approach 
takes into account risk of bias and its relationship to reported effect size
to obtain a conservative and non-negative effect estimate.

In line with the response-surface result, the  significant effect estimate in Brown et al. \cite{Brown19} is accompanied by a low certainty GRADE rating \citep{grade}, in part due to the many high risk of bias studies, along with the caveat that ``the true effect may be substantially different from the estimate of the effect.'' In other words, the authors note in the analysis that the reported effect estimate is questionable. However, reporting a result with a confidence interval while conceding that it may be far from the true effect, although good, is not the same as conducting a response-surface meta-analysis, which seeks to directly estimate this true scientific effect $\tau$ and report an accurate reflection of the uncertainty around the current estimate of this effect. 

\section{Challenges in Response-Surface Meta-Analysis}
\label{sec:discussion}
The simulated and real-world examples presented in the previous sections illustrate why the results of a response-surface meta-analysis can differ from those of literature synthesis meta-analysis. This difference is due to (i) the targeted estimand and (ii) the assumption of how study design quality affects the study-specific estimates. For instance, if an effect becomes more pronounced as the studies get better, then traditional meta-analysis might underestimate the true effect whereas response-surface meta-analysis should yield more accurate estimates by incorporating the relationship between design quality and the effect estimate. Our synthetic setup provided an example of this situation. On the other hand, literature synthesis meta-analysis might claim an effect due the results of lower-quality studies whereas response-surface meta-analysis is more conservative 
by incorporating the
uncertainty that arises when extrapolating to the perfect study. Our real-world case study reflects this.

Although there is conceptual clarity in the response-surface formulation of meta-analysis, there are practical challenges in its implementation. In particular, response-surface meta-analysis can be sensitive to choices about how design quality is scored and how the response-surface is modeled. We believe that the lack of standard guidance along these axes is the key reason why this approach to meta-analysis has not been adopted. Thus, to encourage adoption, we offer initial solutions and suggestions for standardizing the response-surface approach to meta-analysis.

\subsection{Determining Design Quality}
The results of a response-surface meta-analysis depend highly on how design quality scores are assigned to existing studies and tothe idealized study. Without a standard procedure to assign design quality, it is easy to dismiss the results of a response-surface meta-analysis simply by disagreeing with the design quality scale used. For instance, in the above real-data example, we used a coarse rating based on the Cochrane RoB2 such that three studies out of 25 were assigned the same quality as an idealized study. Some may argue that it is inappropriate to assign any existing study a quality score of $z_\infty$, while others may argue against such a coarse assignment. Yet others may advocate for an entirely different scale. In each of these cases, a disagreement with the quality assignment can lead to distrust in the results of the analysis. Thus, for the results of a response-surface meta-analysis to be widely accepted, it is important to establish standard guidelines for assigning design quality.

Finding a standard rating that is generally accepted will require community discussion and engagement. This issue is not specific to the response-surface formulation of meta-analysis, however. 
For instance, one could just as easily disagree with the inclusion/exclusion criteria used in a literature synthesis meta-analysis; existing meta-analyses address potential concerns by defining eligibility criteria in advance and justifying the criteria \cite{McKenzie}, and similar strategies can help legitimize the results of a response-surface meta-analysis.

Tools such as the Cochrane RoB2 tool for randomized control trials \cite{HigginsBias} and ROBINS-I risk of bias tool for observational studies \cite{Sterne} already capture some degree of study quality. However, such tools imply that randomized control trials and observational studies should be treated separately and cannot be ranked using a common set of criteria. 
A response-surface meta-analysis, on the other hand, may wish to take advantage of all studies in estimating a response-surface, which would require a way to rate different studies relative to each other. 

The difficulty in assigning quality scores has been discussed in the past \cite{Greenaland2001,Juni}. 
However, this difficulty should not deter the scientific community from pursuing the estimation of the true estimand of interest. Strategies for incorporating quality scores in a scientific manner include setting community standards for scoring design quality, assigning quality using study design alone without knowledge of reported effect, and conducting sensitivity analyses based on a range of plausible assignments.

\subsection{Modeling the response-surface} 
The results of a response-surface meta-analysis are also generally sensitive to how the relationship between design quality and effect size is modeled.
Put another way, it is not possible to draw inferences about perfect studies using imperfect studies without assumptions on how such study results relate to each other. 

One strategy to help enable accurate response surfaces is to build and incorporate external information about the role of study design and effect size.
For instance, information from historical patterns or field-specific knowledge could help suggest the types of relationships that are plausible. Additionally, the relationships estimated on widely-studied questions could help inform the modeling assumptions of a response surface on newer questions, given a plausible justification for why one would expect these relationships to be similar.
To carry this idea out on a more systematic scale, individual fields of study could also invest in estimating and understanding the causal relationship between study design factors and reported effect size. Such estimates could then benefit a wide array of downstream meta-analyses which could incorporate such prior knowledge into response surface modeling efforts.

Another way to increase confidence in the conclusions of a response-surface meta-analysis is to conduct a sensitivity analysis based on various modeling choices.   
For instance, in the above real-world example, we reported an estimate of 0.01 (95\% CI: -0.17, 0.19) using a random-effects response-surface when encoding risk of bias as an ordinal variable.
If we instead encoded risk of bias as a numeric variable, representing the assumption of a linear relationship between equally spaced risk of bias scores, the resulting prediction mean at $z_\infty$ is also 0.01 (95\% CI: -0.07, 0.09), with a tighter confidence interval due to stronger assumptions on how the different qualities relate. Alternatively, ignoring the high and unclear risk of bias studies all together and simply fitting a random-effects model on the low risk of bias studies also yields a non-negative estimate of 0.00 (95\% CI: -0.04, 0.05). This result has an even tighter confidence band because incorporating high and unclear risk of bias studies introduces additional unexplained heterogeneity in the aforementioned models. Though the uncertainties around these different response-surface variants differ, all the resulting estimates are non-negative and the overall conclusion is the same, that there is not enough evidence to favor the intervention.
Reporting multiple results obtained from different modeling choices can help increase confidence that the conclusions of a meta-analysis are not simply the results of modeling-specific choices. Other strategies to build confidence in a response-surface meta-analysis include reporting several model summaries and evaluations.

\section{Conclusion}
In this work, we reintroduce response-surface meta-analysis, which reframes meta analysis as an endeavor to estimate true scientific effects as would be measured under perfect studies, rather than as a way to summarize the existing population of imperfect studies. Much of the statistical machinery developed for meta-regression can be used to fit a response-surface of the individual studies, thereby enabling an estimate under the perfect study via a predicted mean. We illustrate how the results of a response-surface meta-analysis can differ from traditional approaches. Finally, we discuss 
two main challenges to implementing response-surface meta-analyses. The first is determining a rating system for design quality, which requires consensus building within research communities. The second is modeling and extrapolating from the response-surface, which would benefit from well-formulated assumptions and comprehensive sensitivity analyses. These challenges may seem daunting, but they reflect the effort required to make inferences about true scientific effects in the presence of imperfect data.

Our attitude toward an endeavor such as a scientific meta-analysis is that it must start with a clear statement of the objective of the investigation: namely, the causal estimand of interest. It is important to realize that the underlying estimand targeted by current meta-analysis methods is the typical result among the collection of studies currently conducted by researchers in a particular field. Although a valid question, it is often distinct from the question of the real scientific effect. In some ways, the current literature synthesis approach is analogous to the intention-to-treat (ITT) approach often used in human experiments when there is noncompliance. ``Intention-to-treat'' estimates the causal effects of the assignment to take a drug versus a placebo, ignoring whether a person actually takes the drug or not.  As discussed previously \cite{Sheiner}, this estimation process mixes two different scientific effects: the medical effect of the drug itself, which is the typical objective when conducting the study, and the combined effect of the drug and the participant's willingness to comply with the assignment in the trial setting, a context which differs from the context of \textit{your} doctor treating \textit{you} for a disease. In a similar way, literature synthesis confuses two scientific questions:  the true effect, which would be estimated by a perfect study, and the average effect in the choice of studies conducted by current researchers. Both questions are legitimate, but we believe current practice in meta-analysis usually addresses the second while reporting results as if the first had been addressed. Understanding the differences can lead to a more appropriate interpretation of the results of any meta-analysis. We hope that this article helps to encourage more clarity in research and educational training about meta-analysis, while inspiring future work towards a more 
complete and direct formulation.

\section{Acknowledgements}
Research reported in this publication was supported by the John Harvard Distinguished Science Fellows Program within the FAS Division of Science of Harvard University; by the Office of the Director, National Institutes of Health under Award Numbers DP5OD021412 and R01AI102710; and by NRT-HDR: FUTURE and NSF Award 1922658. The content is solely the responsibility of the authors and does not necessarily represent the official views of the National Institutes of Health or National Science Foundation.

\end{document}